# Blockchain for Energy Market: A Comprehensive Survey

**Tianqi Jiang [a], Haoxiang Luo [b, *], Kun Yang [c], Gang Sun [b, 1, *], Hongfang Yu [b, d], Qi Huang [e, f, 2], Athanasios V. Vasilakos [g, 1]**

[a] Glasgow College, University of Electronic Science and Technology of China, Chengdu 611731, China

[b] Key Laboratory of Optical Fiber Sensing and Communications (Ministry of Education), University of Electronic Science and Technology of China, Chengdu 611731, China

[c] Guangyuan Power Supply Company, State Grid Sichuan Electric Power Company, Guangyuan 628033, China

[d] Peng Cheng Laboratory, Shenzhen 518066, China

[e] Sichuan Provincial Key Laboratory of Power System Wide-Area Measurement and Control, University of Electronic Science and Technology of China, Chengdu 611731, China

[f] School of Information Engineering, Southwest University of Science and Technology, Mianyang 621010, China

[g] Center for AI Research, University of Agder, 4879 Grimstad, Norway

**Abstract**

The energy market encompasses the behavior of energy supply and trading within a platform system. By utilizing centralized or distributed trading, energy can be effectively managed and distributed across different regions, thereby achieving market equilibrium and satisfying both producers and consumers. Additionally, the energy market can address future production control and environmental concerns such as energy shortages and environmental deterioration. However, recent years have presented unprecedented challenges and difficulties for the development of the energy market. These challenges include regional energy imbalances, volatile energy pricing, high computing costs, and issues related to transaction information disclosure. These factors have hindered the smooth operation of the energy market. Researchers widely acknowledge that the security features of blockchain technology can enhance the efficiency of energy transactions and establish the fundamental stability and robustness of the energy market. This type of blockchain-enabled energy market is commonly referred to as an energy blockchain. Currently, there is a burgeoning amount of research in this field, encompassing algorithm design, framework construction, and practical application. It is crucial to organize and compare these research efforts to facilitate the further advancement of energy blockchain.

Tianqi Jiang and Haoxiang Luo have equal contributions to this work.

* Corresponding author. E-mail address: lhx991115@163.com (Haoxiang Luo); gangsun@uestc.edu.cn (Gang Sun)

1. Senior Member of IEEE; 2. Fellow of IEEE

This survey aims to comprehensively review the fundamental characteristics of blockchain and energy markets, highlighting the significant advantages of combining the two. Moreover, based on existing research outcomes, we will categorize and compare the current energy market research supported by blockchain in terms of algorithm design, market framework construction, and the policies and practical applications adopted by different countries. Finally, we will address current issues and propose potential future directions for improvement, to provide guidance for the practical implementation of blockchain in the energy market.

1. Introduction

Energy plays a pivotal role in addressing global climate and environmental challenges, with potential implications for mitigating the impacts of climate change through energy conservation and emission reduction [1]. The World Energy Council (WEC) conducted a comparative analysis of the energy systems in 128 countries, assessing national energy policies and performance across three key dimensions: energy security, energy equity, and environmental sustainability. This evaluation underscores the value of proactive, sustainable energy policies in fostering long-term benefits [2]. The enhancement of the energy market's status is intricately tied to the fundamental importance of energy. Essentially, the energy market operates as a behavioral process facilitating energy trading and supply, encompassing electricity as well as other energy commodities such as natural gas, as it originated from the needs of electricity generation and distribution [3]. Energy policy reform efforts undertaken by various nations are focused on transitioning from conventional to distributed energy markets, as highlighted in research on energy system policies [4]. The connection between policy initiatives and user motivation emphasizes the positive outcomes that well-crafted policies can yield.

The conventional energy market operates within a centralized framework, necessitating a central control unit for energy management and processing. Previous research by [5] and [6] has examined the operational dynamics within this centralized structure. While the centralized model offers advantages, such as enabling the central operator to analyze data and optimize decision-making in alignment with the collective objectives of producers and consumers, it also poses challenges. Specifically, the central control process heightens the risks of information opacity and user data compromise, leading to data complexity and potential bottlenecks that impede informed decision-making,

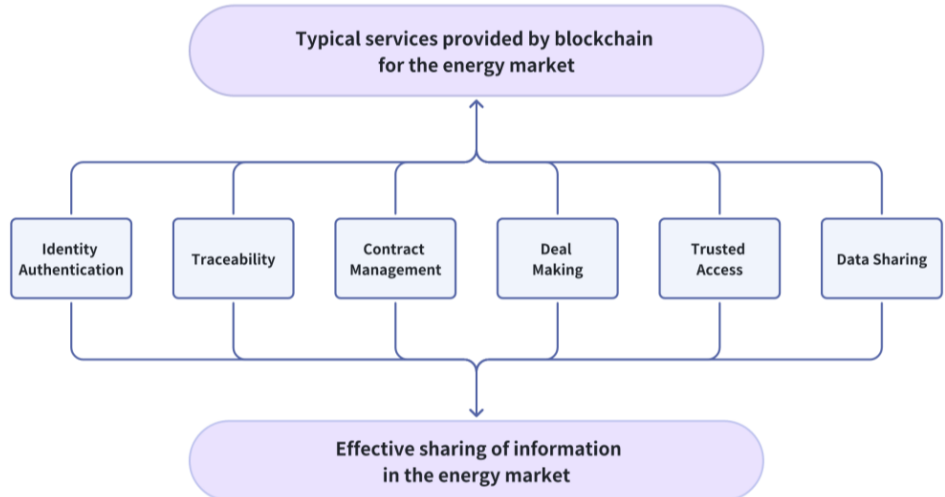

Fig1. Possible applications of blockchain in the energy market ultimately diminishing market efficiency [7-9]. Should the platform's operational efficiency remain stagnant, the integrity of the entire energy market ecosystem is at stake [10]. Furthermore, scholarly investigations like [11] have underscored the exorbitant costs associated with centrally supplied energy compared to distributed energy markets. Moreover, centralization is marked by limited scalability, constraining the geographical reach and adaptability of the energy market. In light of the discernible trajectory of the energy industry's evolution, technologies like blockchain, big data, and artificial intelligence are poised to be integrated deeply into the energy sector [12]. The gradual transition of the energy market from centralization to a distributed model holds significant promise. The transformative potential of blockchain technology in advancing distributed energy markets has been highlighted, as elucidated in Fig 1. For instance, Ridyard et al. [13] have emphasized the blockchain's role in driving the widespread adoption of distributed energy markets, exemplified by innovations like smart microgrids that offer decentralized alternatives to traditional centralized markets.

Over the past few decades, the exponential growth in data storage capacity, processing capabilities, and network computing infrastructure has generated a vast network of interconnected information. This evolution has witnessed a transformation of data from isolated entities to an ecosystem characterized by open online transactions [14]. Blockchain technology, categorized as a form of distributed ledger technology (DLT), comprises interconnected blocks whose integrity and relationships are securely maintained. Initially employed solely for the storage of digital currency transactions, blockchain has since found applicability beyond monetary exchanges and payment processing [15,16]. Viewed through this lens, blockchain transcends its initial identity as solely a financial tool and assumes the characteristics of a general-purpose technology (GPT). Typically adopted by downstream sectors, the innovative and enhanced deployment of GPTs holds the potential to propel economic growth and enhance overall productivity [17,18].

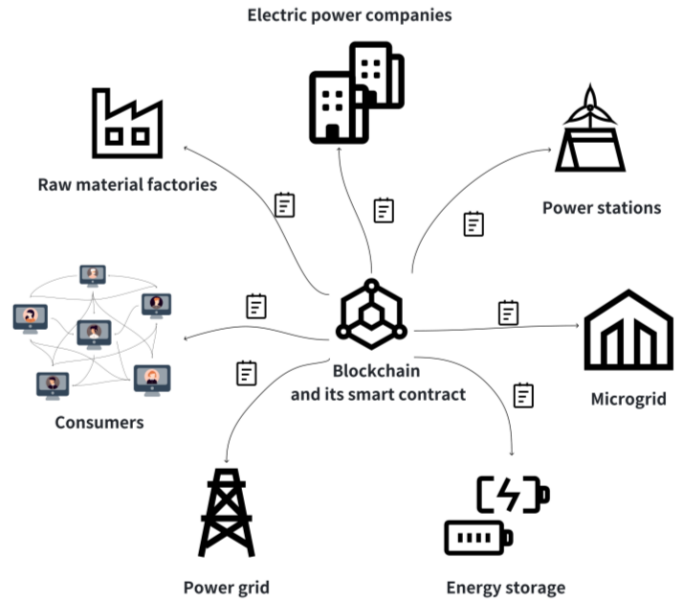

Fig 2. The framework of the blockchain-based energy market.

A significant proliferation of distributed energy producers and consumers has transpired, with prominent entities like large-scale factories deploying their power generation infrastructure. Concurrently, individual stakeholders including households and users have taken initiatives to establish personal energy storage systems, facilitating the sale of surplus energy resources to recoup economic gains. This surge in distributed energy utilization is poised to catalyze a paradigm shift in the energy market, fostering deeper integration with blockchain technology [13]. Fig. 2 delineates the conceptual framework and domain coverage of a blockchain-enabled energy market.

Within the overarching trajectory of industrial digitization and intelligent evolution, the energy sector has progressively transitioned into the digital energy epoch. Presently, a predominant challenge facing many energy practitioners revolves around elevated operational costs and associated inefficiencies. In response, they are turning towards blockchain technology as a transformative tool to forge a new low-carbon economic landscape. This framework aims to address the burgeoning complexity of transactional demands within the increasingly decentralized energy landscape, all while safeguarding energy supply security and accommodating the burgeoning trend of distributed energy systems. The integration of blockchain technology in the energy domain offers the promise of enhancing the agility of energy markets, streamlining operational processes, and streamlining regulatory procedures [19-20]. Functioning as a novel form of distributed infrastructure, blockchain presents a robust and efficient platform for executing and recording energy transactions, anchoring vast troves of secure, immutable data accessible for regulatory scrutiny. A robust energy blockchain framework can offer a gamut of services, encompassing identity verification, certificate tracking, contract administration, transaction facilitation, secure access provision, data sharing mechanisms, among others, fostering effective information exchange among upstream and downstream entities within the energy realm. Illustrated in Fig. 3, its bidirectional mechanism underpins the security fortification of energy markets [21-22].

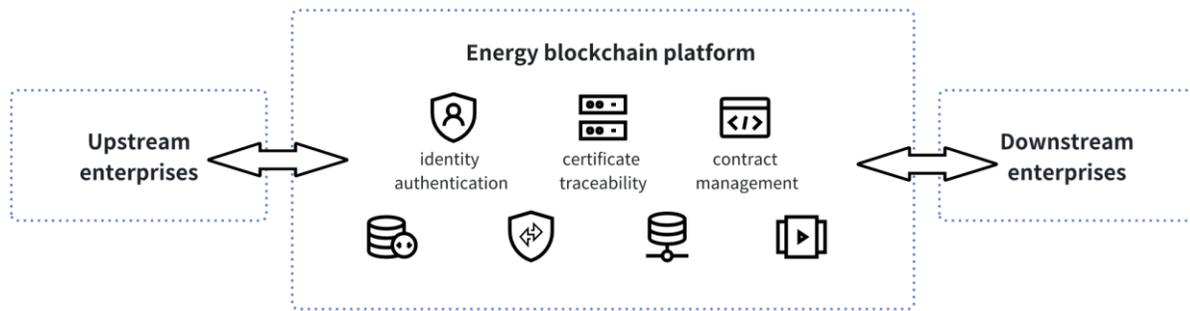

Fig 3. Energy blockchain platform.

Blockchain technology has emerged as a promising solution to address the prevailing centralized bottlenecks within the energy market, prompting researchers to dedicate considerable efforts in investigating the underlying technical intricacies of energy blockchain platforms and associated operational frameworks. This study presents a comprehensive survey encapsulating the spectrum of blockchain-based applications in the energy sector, encompassing an introductory discourse on the evolution of blockchain technology and its intersection with the energy market. Delving deeper, it expounds on the symbiotic relationship between the two domains, elucidating the diverse technical methodologies and applications driving the vibrant developments within this domain. Furthermore, the study delves into establishing a bridge with practical communities through theoretical models, such as comparative analyses of real-world applications across different nations to discern the prevailing development trajectory of blockchain-integrated energy markets. The study endeavors to proffer solutions to the following pertinent challenges encountered in this realm.

- How can blockchain solve the major problems facing the energy market?
- How can the existing energy market framework be improved?

- What efforts have countries made in recent years to enrich and develop the blockchain-enabled energy market? How successful and widespread is it?

Through additional investigation, this study delineates a path forward for inspiring innovation and advancement among future researchers in the field.

**1.1. Contribution**

In the context of broad-ranging literature reviews on blockchain and energy markets, prior studies have diverged in their focus areas, with some emphasizing the evolutionary trajectory and innovations within blockchain technology, others highlighting the outcomes of blockchain applications in energy markets through cross-national policy comparisons, and certain works projecting future developments by analyzing energy trading outcomes. Each of these works prioritizes the advancement of energy blockchain in distinct dimensions. Yet, this study contends that the integration of blockchain in the energy market warrants not only a granular examination from technical and applied perspectives but also necessitates an exploration of its implications on energy storage, distribution, and trading dynamics. Furthermore, we propose that the nuanced application of blockchain technology in specific contexts and simulated scenarios can yield superior outcomes. Our research outcomes contribute novel insights by bridging theory and practice, expanding the horizons of existing literature reviews. By amalgamating technological insights with market evaluations, we aim to enhance the reliability and efficacy of the comprehensive survey.

The contributions of our survey mainly lie in the following aspects.

- As a survey describing the current evolution of blockchain-based energy markets, we have a comprehensive look at why blockchains are suitable for the energy market, as well as the current technical bottlenecks of blockchains.
- By synthesizing pertinent literature reviews authored by prominent experts in the domain, we undertake a rigorous analysis of blockchain applications in the energy market, encompassing both technological intricacies and practical implementations.
- In the prospective outlook segment, we proffer solutions to address prevalent challenges encountered in the current landscape, thereby offering guidance on future research directions.

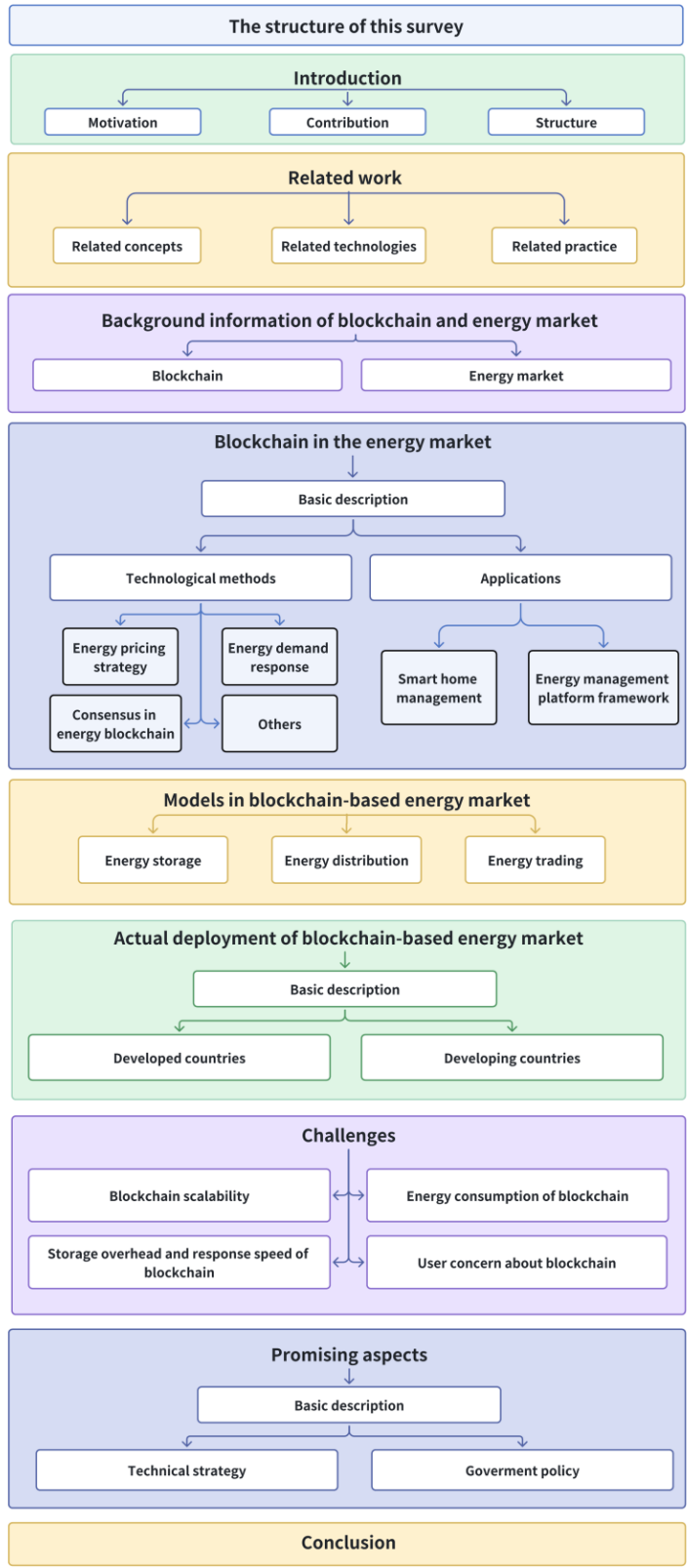

Fig 4. Structure of our survey

Table 1. Summary of existing surveys and their primary focus

| Reference | Concepts of blockchain | Concepts of the energy market | Related technologies | Models and applications | Practical applications | Governments and countries | Future analysis |
|---|---|---|---|---|---|---|---|
| [23] | √ | √ | √ | | | | |
| [24] | | √ | | | | | |
| [25] | | √ | | | | | |
| [26] | | √ | | | | | |
| [27] | | | √ | √ | | | |
| [28] | | √ | √ | √ | | | |
| [29] | | | √ | √ | | | |
| [30] | √ | √ | √ | √ | | √ | |
| [31] | √ | √ | √ | | √ | | |
| [32] | | √ | √ | √ | | | |
| [33] | √ | | √ | √ | | | √ |
| [34] | √ | √ | √ | √ | | | |
| [35] | | √ | √ | √ | | | |
| [36] | | √ | √ | √ | √ | | |
| [37] | | √ | √ | √ | | | |
| [38] | √ | √ | √ | | | | |
| [39] | | √ | √ | √ | | √ | √ |
| [40] | √ | √ | | | √ | √ | √ |
| [41] | √ | | √ | √ | | √ | √ |
| [42] | √ | √ | √ | √ | | | √ |
| [43] | √ | √ | √ | √ | | | √ |
| [44] | | √ | | √ | | √ | √ |
| Our survey | √ | √ | √ | √ | √ | √ | √ |

**1.2. Structure of this survey**

This survey initiates an exploration of blockchain applications in the energy market, delving into the technical algorithmic analysis at the micro-level and the application framework at the macro-level. It delves not only into theoretical models in a regional context but also scrutinizes the practical deployment of blockchain technology in both developed and developing nations globally.

The structure of the survey, as depicted in Fig. 4, unfolds as follows: Section 1 serves as the introduction, following which Section 2 focuses on related works, evaluating 22 reviews with a keen eye on their contributions and limitations. Section 3 furnishes essential background information encompassing the fundamental concepts and characteristics of blockchain and energy markets. In Section 4, the integration of blockchain with the energy market is expounded upon, underscoring its catalytic potential and the transformative impact it engenders. Section 5 navigates through the models underpinning blockchain-driven energy markets, spanning energy storage, distribution, and trading implementations. Moreover, Section 6 spotlights the energy blockchain landscape in key countries worldwide in recent years. Section 7 delves into challenge analysis, delineating the manifold obstacles confronting blockchain applications within the energy market. Lastly, Section 8 foregrounds future predictions and prospects, paving the way for Section 9, which orchestrates the conclusion of this comprehensive survey.

## 2. Related work

In this section, we conduct a classification analysis of prior surveys concerning blockchain-based energy markets, emphasizing the significant impact and contributions of 22 surveys on the comprehensive landscape of blockchain-enabled energy markets. The juxtaposition of these surveys is delineated in Table 1. Our findings reveal a predominant focus in existing literature on concepts, technologies, and applications within this domain. Conceptual discussions serve to elucidate the fundamental definition of entities, constituting the bedrock of understanding. Technological discourse underscores the refined methodologies employed to navigate the evolution and advancements of entities. Application-oriented discussions shed light on the operational functionality of such entities. Hence, this section is dedicated to investigations that expound on conceptual, technological, and practical dimensions of blockchain-based energy markets.

### 2.1 Related concepts

In recent years, a myriad of projects and research endeavors have unfolded within the realm of blockchain-based energy markets. In 2017, Goranovic et al. [23] introduced the pioneering development of microgrids leveraging blockchain technology, expounding on technical frameworks and practical applications. Subsequently, Huang [24] conducted a comprehensive review elucidating the concept and deployment nuances of blockchain-based energy trading systems. Concurrently, Panda et al. [25] navigated the challenges and implications underlying the shift from centralized to decentralized control mechanisms, encapsulating the evolutionary trajectory of the energy market across varying stages. Hirsch et al. [26] predominantly delved into the analysis of pertinent literature and pivotal aspects of microgrids within the energy market domain. In stark contrast to conventional energy markets, the blockchain-based energy market epitomizes a novel application of blockchain technology within the energy landscape, manifesting a fully decentralized energy ecosystem facilitating direct communication between energy producers and consumers for supply-demand transactions. Additionally, blockchain endows consumers with a heightened level of autonomy to engage in direct energy transactions. The scholarly contributions manifested in the aforementioned studies expound upon foundational concepts relevant to blockchain-enabled energy markets. Notably, select academics have accentuated the juxtaposition of energy market reliance on blockchain through a focused technological lens.

## 2.2 Related technologies

Numerous scholars have explored the deployment of pertinent technologies across diverse scenarios. In [27], researchers elucidated the potential of blockchain in enhancing the evolution of the energy market as a future technology. In [28], the authors delved into the intricacies of peer-to-peer (P2P) trading aggregators, distributed energy system (DES) owners, and power grids within the energy market landscape, aiming to optimize stakeholder benefits encompassing economic and temporal gains. Noteworthy iterations of related studies include [29], where Konneh et al dissected the characteristics, merits, and demerits of model predictive control technology and its utility in configuring elastic power systems. Subsequently, [30] scrutinized the clearance methods employed in energy trading markets, examining strategies such as auction theory and game theory. Zou et al. [31] probed the comprehensive integration of energy and blockchain to envisage its transformative impact on energy markets, with a primary objective of delineating viable technological pathways. Furthermore, a myriad of detailed technical analyses are expounded in [32-37], which can be referenced in Table 1 for a consolidated overview.

However, the extant literature predominantly overlooks the disparity between theoretical models and practical implementation. Furthermore, the limitations highlighted underscore that advanced techniques may encounter computational challenges, particularly concerning slow processing speeds when handling extensive community networks. Given the pressing nature of energy market transformation, there arises an urgent need to expedite the application of cutting-edge technologies like blockchain in real-world industrial settings. Our study undertakes a comprehensive analysis of these challenges and proffers preliminary solutions to address them.

## 2.3 Related applications

In addition to a technological dimension, the pertinent academic literature primarily scrutinizes commercial models and societal dynamics. Notably, P2P transactions feature prominently in the discourse surrounding energy sharing and trading across residential, commercial, and industrial settings, with a specific emphasis on business model analyses elucidating the supplier's role [38]. In the year 2022, Schwidtal et al. [39] systematically reviewed the burgeoning business models within the energy market paradigm, juxtaposing the nuances among P2P, community self-consumption, and transactional energy models. Similarly, [40] delved into potential standards, scalability considerations, and other requisite factors in crafting electricity market design frameworks.

Contrasting with the prior studies, Stallone et al. [41], in their 2021 endeavor, delved into the ramifications of blockchain integration on marketing technologies across a cohort of 800 companies, delineating the transformative societal impacts and global applicability of blockchain projects. Additionally, in [42], researchers evaluated blockchain's pivotal role in augmenting the governance dynamics at the nexus of socio-technological energy transformations to fortify energy systems' reliability. These investigations underscore the necessity of blockchain technology in resolving transactional complexities within energy markets.

Afzal et al. [43] outlined an analysis of energy blockchain application and optimization processes, albeit without delving into policy frameworks, practical deployment scenarios, and energy input-output dynamics across diverse nations. Furthermore, [44] delved into the repercussions of distributed energy applications on institutional progress. While these studies provide a nuanced synthesis of such applications, they stop short of offering detailed country-specific exemplars. Contrarily, our research embarks on a comparative analysis by leveraging illustrative examples. This study focalizes on the contemporary advancement and implementation of blockchain-based energy markets across global contexts, positing the challenges encountered by such energy markets and charting plausible solutions and future prospects. The research outcomes hold strategic significance, furnishing pragmatic guidance for stakeholders operating within the blockchain-enabled energy market landscape.

3. **Background knowledge**

The exponential surge in data volume and information dissemination has engendered multifaceted challenges pertaining to trust, privacy, and security. Notably, even solutions derived from big data processing may manifest unreliability [45]. The advent of blockchain technology has emerged as a beacon of hope for numerous researchers aiming to alleviate the regulatory strains afflicting the internet landscape. The architectural blueprint of blockchain infrastructure is delineated across distinct layers encompassing the application layer, contract layer, consensus layer, network layer, and data layer [46]. Noteworthy studies, such as Edelman's Trust Barometer, underscore the imperative to democratize trust, positioning blockchain as an instrumental tool in ushering a transparent and open trust paradigm accessible to the public, aligning with contemporary societal requisites [47]. Beyond elucidating the foundational technology underpinning blockchain, this study aims to illuminate the intricate dynamics of the energy market.

Table 2. Comparison of three blockchain types

| Type of blockchain | Definition | Characteristics | Application | Carrying capacity |
|---|---|---|---|---|
| Public chain | A public, or permission-less, blockchain network is one where anyone can participate without restrictions. Most types of cryptocurrencies run on a public blockchain that is governed by rules or consensus algorithms. | 1. Proof-like consensus mechanism; 2. Anyone to come and go freely; 3. Need incentive mechanism | Virtual currency | Low |
| Consortium chain | A blockchain network where the consensus process is closely controlled by a preselected set of nodes or by a preselected number of stakeholders. | 1. BFT consensus mechanism; 2. Only alliance members to participate in the transaction; 3. Optional authentication mechanisms | Payment; Settlement | Middle |
| Private chain | A private, or permissioned, blockchain allows organizations to set controls on who can access blockchain data. Only users who are granted permission can access specific sets of data. | 1. CFT consensus mechanism; 2. Only individuals or institutions to participate in the transaction 3 No Authentication Mechanism is needed | Audit issue | High |

### 3.1 Blockchain

As early as 2008, Nakamoto [48] introduced the groundbreaking concept of blockchain, heralding a paradigm shift in the economic domain. By obviating the necessity for centralized storage and control entities, blockchain has catalyzed pivotal transformations in contractual agreements, transactions, and record-keeping processes, thereby reshaping the operational blueprint of businesses and industries. Tripathi et al. [49] further underscored the versatile applications of blockchain across diverse sectors, including decision analysis and data processing. The unparalleled speed of updates inherent in blockchain technology has fuelled the exponential growth of the cryptocurrency market.

Blockchain technology serves as a compelling tool for companies to assert security claims, engendering trust via technical means to assuage external stakeholders of the company's ethical conduct. Through embedding business operations and data within blockchain's trust mechanism, enterprises can foster transparency, enabling all pertinent stakeholders to oversee and share critical information, thereby endowing users with rights including access to information, decision-making autonomy, and effective communication channels. This elevation of credibility not only enhances the company's standing but also aligns with the exigencies of end-users and regulatory bodies [50].

Functioning as an advanced distributed ledger, blockchain operates as a shared database utilized for decentralized replication and synchronization among various network nodes [51]. Amidst this operational framework, ensuring the trustworthiness of all uploaded data poses a challenge. Consequently, blockchain not only embodies the salient features of a distributed ledger but also integrates consensus algorithms and smart contracts to mitigate the presence of malicious nodes and uphold the integrity of uploaded information. In a systematic evaluation presented in [52], the authors expounded on blockchain's modular structure, delving into key aspects such as definition, classification, and application scenarios.

Blockchain architectures are categorized into public, consortium, and private chains based on the extent of network propagation, each offering distinct advantages. Despite their merits, variations in practical applications, user involvement, and processing speeds necessitate careful consideration [53-55]. The public chain epitomizes complete decentralization, allowing nodes to join or exit the network at will. In contrast, the consortium chain exhibits partial decentralization managed by multiple entities, requiring user authentication for network entry. Indicatively, the private chain resembles conventional databases and poses limitations for large-scale distributed energy markets.

Historically, early blockchain-based energy markets predominantly leveraged public chains [56] due to the sector's extensive involvement of electricity producers and consumers. Nonetheless, with advancements in consensus mechanisms, energy markets have shifted towards adopting consortium chains [51]. The consortium chain enforces admission protocols, ensuring participant compliance, enhancing data privacy through permission controls, and accelerating consensus efficiency via Byzantine fault-tolerant (BFT) algorithms. Noteworthy also is the controllability of node volume in the consortium chain, coupled with the flexibility in employing swifter consensus mechanisms.

The application landscapes of blockchain extend across digital medical services, green and renewable energy sectors, smart homes, intelligent transportation systems, and financial operations [60-61]. As illustrated in Fig. 5, blockchain's application horizons span diverse realms, charting a course for innovation and efficiency in these sectors.

The advantages of blockchain technology are readily apparent [62-63], underscored by its (1) robust security measures: leveraging a distributed ledger framework, each node maintains an identical and comprehensive ledger, ensuring that the compromise of a single node does not compromise the integrity of the entire ledger and transaction records; (2) heightened transparency: alongside encrypting the private information of transaction participants, blockchain enables users to access data and develop tailored applications through open interfaces, fostering a transparent ecosystem; (3) immutability: once information is verified and added to the blockchain, it becomes permanently stored, constituting a chronologically labeled data record resistant to unauthorized alterations. This feature not only guarantees data traceability but also significantly diminishes the risk of data manipulation by various nodes at any given time. These distinctive advantages hold profound value across the specified application domains.

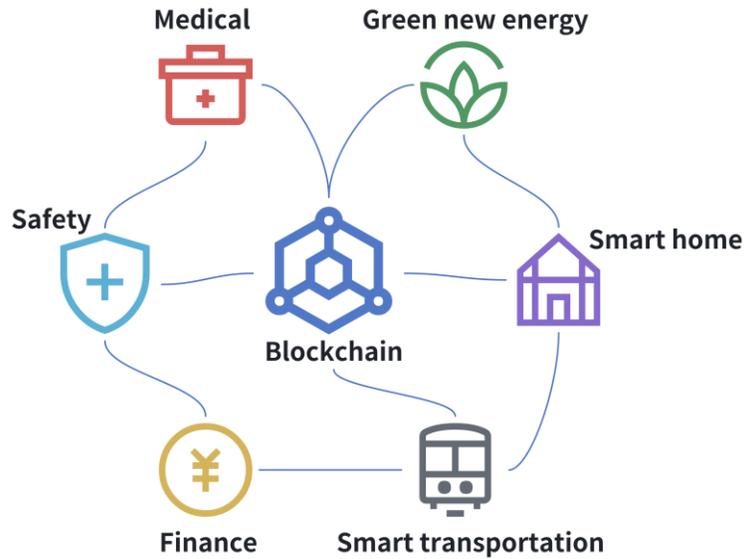

Fig 5. Application scenarios of blockchain

**3.2 Energy market**

Electric power plays a significant role within the energy market landscape [64], particularly within the realm of renewable energy systems. Hence, our research primarily centers on electricity supply and demand dynamics within the energy market domain. A critical challenge lies in ensuring the reliability and flexibility of transactions within the energy market. In [65], the necessity for a robust technical infrastructure to support an autonomous trading platform was underscored, stressing that the current state of energy market development falls short of achieving comprehensive regional coverage.

Delving into the rationale behind the evolution of energy markets unveils a myriad of advantages. Investors recognize that the nascent energy market holds the promise of stabilizing energy transitions, as evidenced by the insights shared in [66]. Concurrently, in [67], the expansive potential of energy markets across diverse regions, including neighborhood and community applications, was highlighted. Wiyono et al. [68] elucidated the significance of energy markets through the efficient utilization of surplus power, advocating for intelligent management of distributed community energy to enhance overall utility. Furthermore, Pavic et al. [69] deliberated on maximizing benefits by juxtaposing operational frameworks of distributed energy markets in developed nations.

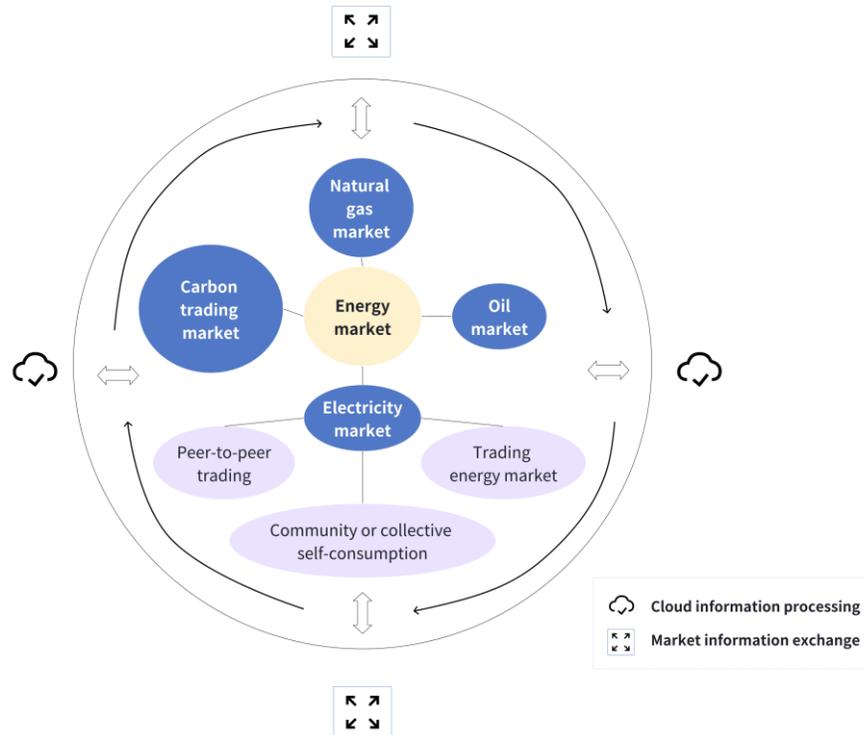

Fig 6. Complex energy markets

Furthermore, as attention progressively shifts from traditional to renewable energy sources like photovoltaic and wind power generation, the necessity for energy markets intensifies. Notably, the burgeoning focus on distributed new energy sources underscores this demand for energy markets. In [70], the assertion that "humanity is undergoing an energy structural transformation" underscores the profound impact on national development trajectories. Consequently, the energy market emerges as a pivotal public infrastructure dictating energy flow and distribution patterns, encapsulating significant research relevance.

Nevertheless, the complexity surrounding information and trading systems within the energy market, as depicted in Fig. 6, encompassing the electricity, oil, natural gas, and carbon trading markets [71-72], underscores inherent challenges. Moreover, the integration of distributed power energy through local energy markets, which predominate the electricity market landscape, introduces distinctive operational modalities. Notably, P2P trading, community or collective self-consumption, and trading energy markets represent prevalent modes within local energy markets, each characterized by differing operational scales and trade objectives [39]. Presently, existing market regulatory frameworks prove inadequate to support the comprehensive operations of the energy sector. Consequently, shifting attention towards blockchain emerges as a plausible strategy to address the inefficiencies in financial and material resource allocation compared to traditional internet technologies [73].

## 4. Blockchain in the energy market

As outlined in [74], distributed energy markets encompass application scenarios at the small building, district, and urban levels. Blockchain-supported energy markets primarily include grid power generation, electricity consumption, demand response for smart homes, and electric vehicles. This study delves into the demand response of grid generation and electricity consumption, as well as smart home integration.

Blockchain technology, as an emerging Information and Communication Technology (ICT), assumes various essential functions within the energy market [75]. By eliminating the need for intermediaries in transactions, blockchain offers a potential avenue for balancing energy demand across diverse regions [76]. Moreover, the increasing interest in blockchain for facilitating secure transactions can expedite auditing processes, albeit it does not supplant human supervision entirely. Auditors must possess a heightened level of expertise to assess transaction ledger accuracy and adeptly address potential data discrepancies [77-78].

Emphasizing the power generation and consumption domains, the blockchain-based energy market swiftly delivers clean energy to users, fostering climate change mitigation efforts through supply-side innovation. The fundamental utility lies in the traceability and decentralization mechanisms inherent in blockchain technology [79]. Amidst energy homogenization challenges, blockchain introduces competitive advantages for power companies, including product differentiation through diversified offerings, data transparency, and auditability, and the establishment of open and trusted operational protocols benefiting both suppliers and users [80-81].

In the short term, blockchain may impose constraints on power companies by diminishing centralized authority. However, over the long term, the security assurances engendered by blockchain instill confidence and satisfaction among users, facilitating market penetration. Furthermore, blockchain solutions hold promise for alleviating market pain points stemming from energy commoditization [80-81].

### 4.1 Technical methods

Technical means primarily entail the development of novel methodologies or algorithms aimed at enhancing specific performance metrics. This section emphasizes the optimization of algorithms and methods within the energy blockchain domain to maximize utility in the energy market, encompassing facets such as energy pricing strategies, energy demand response mechanisms, and blockchain consensus algorithms.

**4.1.1 Energy pricing strategy**

Given that energy trading constitutes a significant segment of the energy market, with predetermined pricing conditions underpinning various transactions, the development of effective pricing strategies and decision-making formulas holds paramount importance and necessitates continual enhancement. A double auction method stands out as a prevalent approach [82], adept at optimizing utility for both buyers and sellers. Game theory serves as a valuable modeling tool utilized by researchers to address the intricate balance between energy production and demand [83-85], ultimately enabling a comprehensive understanding of the stakeholders involved and the associated payoffs [86-87]. Notably, power producers, consumers, and interactions among various consumer groups may engage in strategic gameplay dynamics [86].

Researchers are actively exploring avenues to streamline energy management processes, thereby reducing transaction costs to attract a broader customer base. For instance, the model proposed in [88] offers comprehensive management of hourly and daily energy consumption for retailers, facilitating cost reduction and enhancing pricing strategies to benefit end-users. Concurrently, initiatives like the blockchain-based clearing platform introduced by Hamouda et al. [89] pave the way for energy trading in interconnected microgrids, fostering operational efficiency and cost optimization through modular and flexible energy transactions. Additionally, the integration of batch call functions for contract creation, as depicted in [90], bolsters trust and user reliance. Moreover, innovative decentralized sorting algorithms have been designed to empower the lowest bidders to sell surplus energy and maximize their returns [91].

Moreover, recognizing energy as a key data production factor [92], the energy pricing mechanism must factor in three key considerations: (1) scenario-specific value attribution necessitating situation-based pricing strategies; (2) the interplay between bilateral and multilateral market dynamics influencing pricing frameworks; (3) the nuanced nature of energy trading, wherein contracts can pertain to either the right to use or ownership, thus mandating tailored contract designs to align with distinct trading rights.

Evidently, energy pricing strategies are not solely determined by pricing methodologies, but are also intricately influenced by the institutional arrangements orchestrating interactions between buyers and sellers within the framework of varying constraints such as systemic, scenario-specific, and technological parameters. Detailed analysis of specific scenarios shaping energy pricing strategies will be elucidated in Section 5.

**4.1.2 Energy demand response**

The concept of demand response was officially delineated by the United States Federal Energy Regulatory Commission in 2010 [93], distinguishing itself from pricing strategies primarily governed by power producers. Demand response predominantly revolves around the proactive actions of power consumers, entailing the temporary adjustment of energy consumption patterns in response to price signals or incentives. This adaptive behavior serves to curtail electricity usage or facilitate its redistribution for a stipulated period, thereby upholding power system stability and averting surges in electricity pricing.

Integral to the efficacy of demand response is the user's experiential perception, encapsulated by satisfaction surveys. Numerous scholars leverage comparative analysis between anticipated and actual responses to refine estimations, thereby enhancing user sensibility toward energy prices. Notably, in 2017, Chen et al. proposed a framework transitioning from demand response to interactive energy, fostering a harmonious equilibrium between supply and demand through flexible distributed load power generation resources. This approach unearthed the fundamental principles underpinning demand response and interactive energy, centering on real-time optimization within an interactive distribution system to augment process flexibility and optimize individual welfare through information exchange among stakeholders [94-95]. Such a strategy not only prioritizes demand-side considerations but also underscores the significance of user satisfaction.

The United States allocates considerable attention to this trajectory, foreseeing practical challenges in demand response evolution, such as passive user involvement, the absence of standardized demand response pricing protocols, and diminishing cost-effectiveness [96]. Failure to engage users satisfactorily in energy demand response may impinge upon their interests, thereby impeding the advancement of blockchain-integrated energy markets. To mitigate these challenges, a meticulously crafted blockchain consensus mechanism emerges as a judicious approach to fostering user engagement in demand response initiatives [97], with detailed consensus frameworks addressed in Section 4.1.3.

Moreover, insights shared in [98] shed light on demand-side management (DSM) within diverse distributed power markets, advocating for tailored deployment strategies contingent upon each country's specific energy generation mix. DSM not only enhances system activity and reliability as demand response efficacy improves but also tempers response agility. Additionally, a proposed optimization strategy for demand response in [99] not only resolves economic quandaries but also ensures energy price stability over a short period, curtailing market volatility.

**4.1.3 Consensus in energy blockchain**

In contrast to the preceding parts focusing on the energy market domain, the current segment delves into the foundational technology of blockchain: the consensus algorithm. Consensus mechanisms play a pivotal role in fostering alignment among entities within the energy market ecosystem sans the intervention of third-party trust entities, serving as a critical enabler for decentralizing the energy market [51]. Within a collaborative network setting, unanimous agreement on actions and events - such as transaction records among members - is imperative for seamless collaboration, with non-conforming entities potentially opting to disengage and terminate cooperation [97].

Historically, the Proof of Work (PoW) consensus algorithm, pioneered by Bitcoin [48], displayed shortcomings in its intricate computational processes, energy-intensive nature, and limited scalability, underscoring the exigency for novel consensus algorithm explorations. Notably, prevalent consensus algorithms, as outlined in [100-101], often present drawbacks impeding blockchain advancement. To enhance blockchain's utility in serving the energy market, pivotal areas for improvement include consensus speed, throughput, and security considerations [51, 97], aiming to enhance operational efficiency, transaction scalability, and system robustness.

In the pursuit of enhancing consensus mechanisms, two primary avenues emerge: the superposition consensus approach and the exploration of novel consensus frameworks. Superposition consensus amalgamates the strengths of multiple consensus algorithms to forge a new consensus model, exemplified by mechanisms like Proof of Capacity (PoC) [102], which melds the performance benefits of Delegated Proof-of-Stake Consensus (DPoS) while mitigating inherent centralization risks through credit voting mechanisms. Moreover, developments such as PRAFT (Practical Byzantine Fault Tolerance, PBFT-enabled RAFT) and RPBFT (RAFT-enabled PBFT) epitomize novel consensus models combining the advantages of PBFT and RAFT, enhancing security and consensus efficiency over their predecessors [51].

Furthermore, innovative consensus models tailored for energy markets have emerged, exemplified by the ByzCoinX BFT consensus protocol fortifying market resilience against DoS attacks [104]. Additionally, relaxed-consensus models proposed by Sorin et al. [105] innovate decentralized approaches to addressing scheduling challenges in distributed energy markets, while Proof-of-Computational Closeness (PoCC) and Practical Proof of

Table 3. Consensus in energy blockchain

| Reference | Consensus name | Design concept | Characteristic | Applied model |
|---|---|---|---|---|
| [51] | PBFT-enabled RAFT (PRAFT) and RAFT-enabled PBFT (RPBFT) | Combining the advantages of PBFT and RAFT | Low complexity, tolerant of Byzantine nodes | Energy trading |
| [97] | Proof of Solution (PoSo) | Using mathematical optimization methods to solve energy-consuming mining problems in PoW | Easy to verify consensus results without wasting computing power | Energy distribution and trading |
| [102] | Proof of Capacity (PoC) | Using the hard disk storage space as a competitor condition | Low cost, low energy consumption | Energy storage |
| [105] | Relaxed Consensus + Innovation (RCI) | Resemble the dual ascent method, splitting the global optimization problem into small local problems | Completely decentralized | Energy distribution |
| [106] | Proof-of-Computational Closeness (PoCC) | For the selection of miners and the creation of blocks | Highly scalable and resistant to Sybil attacks | Energy trading |

Storage (PPoS) mechanisms introduced by Samuel et al. [106] and [107] respectively demonstrate robust security enhancements leveraging advanced encryption architectures. Notably, the transformative work by Chen et al. reimagines PoW mining challenges, presenting a novel Proof of Solution (PoSo) consensus algorithm tailored for energy scheduling within the energy market, optimizing verification costs and enhancing Byzantine fault tolerance capabilities [97,108]. The detailed applications of blockchain consensus in the energy market context are succinctly synthesized in Table 3.

**4.1.4 Others**

Cryptography plays a pivotal role in bolstering the security infrastructure of blockchain technology, thus enhancing its reliability and resilience in catering to the energy market [109]. For instance, the integration of blockchain and zero-knowledge proof offers a framework for implementing privacy-centric energy trading mechanisms [7]. Furthermore, the application of artificial intelligence, particularly reinforcement learning (RL), demonstrates superior computational efficiency in optimizing power dispatch operations. RL adeptly navigates uncertain energy distributions to facilitate seamless online P2P information exchanges, maximizing desired outcomes and positioning itself as a key enabler for resolving future energy market constraints [109]. Concurrently, methodologies such as Lyapunov optimization technology, alternate direction methods, and multi-objective optimization approaches are tailored toward streamlining system operation costs [110-112].

These refined technical strategies collectively tackle the distributed developmental trajectory of blockchain-based energy markets, presenting foundational solutions from an academic standpoint.

## 4.2 Applications

The term "application" in this context pertains to the deployment of innovative solutions within blockchain-based energy markets, emphasizing practical innovations instead of theoretical algorithmic advancements. This section prioritizes real-world scenarios over technical processes, aiming to provide a comprehensive view that integrates technical advancements with practical applications in guiding the development of future energy blockchain platforms.

The primary focus of this section centers on the conceptualization and design of an energy management framework. Leveraging blockchain technology, the energy market has transitioned towards evolving into an effective energy management system tailored for microgrids [113]. Positioned as an autonomous energy distribution network and a fundamental element of smart grids, microgrids represent a promising blueprint for advancing future power grid architectures [114]. Therefore, the discussion is initiated by addressing the energy landscape within smart homes, a quintessential illustration of a small-scale microgrid. Subsequently, attention shifts towards exploring energy management platforms designed specifically for microgrids.

### 4.2.1 Smart home management

The concept of a smart home in this context does not center around collective interactions between homes, but rather emphasizes the development of platforms facilitating individualized processes such as P2P energy transactions within smart homes [115], energy transaction models [116], and the microgrid infrastructure within the home [117]. These tailored solutions serve to address energy consumption challenges within small communities through the utilization of energy blockchain platforms. For instance, the energy consumption prediction services enabled by blockchain technology at this stage can yield energy savings, while enhancing community comfort and satisfaction through precise forecasting. Solutions to these challenges typically involve the integration of various layers of existing blockchain frameworks. For instance, in [118], researchers proposed a hybrid energy blockchain system architecture incorporating blockchain smart contracts, decentralized databases, and client interfaces, employing diverse blockchain layers for processing distinct transactional content. Related studies have developed auxiliary frameworks leveraging blockchain technology to reduce error rates and fortify the resilience of energy markets. In [119], researchers reviewed the automation framework of smart homes, assessing the suitability of blockchain for such scenarios. However, findings revealed significant computational challenges impeding widespread blockchain application across all device-generated transactions, necessitating further improvements as elaborated in Section 7.

Furthermore, efforts are directed towards optimizing operational costs within blockchain-enabled smart homes. In [120], researchers conducted simulations of home energy management systems to deliver economic benefits to communities and end-users, aiming to minimize electricity costs. Similarly, Strepparava et al. [121] undertook an analysis of local energy communities to derive optimal solutions within blockchain-based smart homes. Additionally, a segment of housing in Sydney was examined to validate the precision of a building energy management system founded on optimized scheduling and bidding strategies [122]. Notably, the optimal solutions outlined above are grounded in the economic performance of smart home energy transactions.

**4.2.2 Energy management platform**

Energy management platforms, or frameworks, represent proposed methodologies aimed at reconciling energy scheduling and demand response considerations by addressing prevalent energy management challenges and deficiencies, notably computational and time costs. In 2021, Hamouda et al. [123] delineated an energy market framework stratified into blockchain and power system layers, incorporating user marginal pricing considerations while aligning with essential power system requirements. Additionally, in [124], researchers introduced a framework geared towards enhancing the optimal operation of hybrid renewable power plants, mitigating renewable energy volatility, and fostering informed energy management and predictive utilization.

Moreover, artificial intelligence applications harness historical data encompassing energy production and electricity consumption for predictive analytics, harmonizing consumption strategies between energy stakeholders to orchestrate energy market scheduling [125]. However, prevailing power plant operational frameworks are deemed rudimentary, epitomized by simplistic functionalities reliant on machine learning algorithms to bolster power prediction accuracy, consequently lacking robust safety certifications for data learning and predictive outcomes. Consequently, challenges arise in the global green electricity certification and trading framework, hindering result recognition. For instance, the P2P trading framework expounded in [126] not only facilitates conventional energy trading but also encompasses carbon trading, encapsulating diverse energy market transactions necessitating stringent transaction data security validation.

Blockchain technology emerges as a viable solution to address these imperatives, offering intrinsic security features for these platforms. In [127], researchers proposed an energy management framework anchored in blockchain technology, addressing distributed energy economic scheduling challenges through smart contracts to optimize consumer benefits. Subsequently, Dang et al. [128] devised solutions to optimize demand load management problems using blockchain to economize under novel market structures. Furthermore, in [129], authors introduced a blockchain-enabled trading platform catering to energy trading needs, while ensuring cybersecurity requisites. Concurrently, Cali et al. [130] introduced a combined privacy and cybersecurity framework tailored for local energy markets to bolster cyber-physical resilience, spotlighting blockchain as a prospective remedy. The amalgamated research underscores blockchain's potential as an instrumental technology ensuring the cybersecurity of energy market platforms.

## 5. Models in blockchain-based energy market

From an energy market modeling perspective, three prevalent types emerge: energy storage, energy distribution, and energy trading, all predicated on energy management principles [130]. These models primarily rely on simulations involving small-scale users or household communities, utilizing authentic data for accurate analysis and representation.

### 5.1 Energy storage

Energy storage plays a pivotal role in the energy market, particularly within the electricity power system, where it contributes to crucial auxiliary functions such as frequency modulation and peak regulation [131,132]. As early as 2004, Barton et al. [133] delved into the analysis of energy storage, shedding light on its significance in renewable energy application scenarios. Unlike traditional energy sources, renewable energy is characterized by inherent variability and unpredictability, leading to substantial market price fluctuations during off-peak consumption hours. Consequently, the efficient transfer of energy from low-demand to high-demand periods becomes imperative [134]. Energy storage technologies, including air storage, pumped storage, and battery storage, offer viable solutions for power peak regulation by temporarily storing surplus electricity for utilization during periods of heightened demand.

The integration of energy storage technology into energy production and consumption practices has emerged as a key consideration in the low-carbon energy market. While efforts have been directed towards battery capacity enhancement [135] and consumer collaborations [136], the entry of energy storage entities has introduced new dynamics into the energy market landscape, necessitating a reassessment of information security technologies encompassing identity authentication, privacy safeguards, and transaction traceability. Furthermore, stored electrical energy can be traded as a fungible commodity, mirroring concepts akin to bike-sharing programs. Nevertheless, the absence of a reliable transaction and sharing mechanism has hindered the broader adoption of energy storage technologies, underscoring the potential for utilizing blockchain as a robust solution within the energy market ecosystem.

Ongoing research endeavors are actively exploring the convergence of energy storage and blockchain technology. For instance, in [137], researchers highlighted the viability of shared energy storage at community levels, positioning traditional grid networks as forms of large-scale storage infrastructure. Additionally, in [138], Luo et al. proposed an energy storage sharing mechanism within private blockchain networks to facilitate secure and transparent transactions in joint markets. Furthermore, experiments conducted by Niklas et al. [139] demonstrated the capacity of energy storage systems to effectively integrate fluctuating power generation outputs and electricity demand, provided that their recognized value and services align with prevailing market structures.

**5.2 Energy distribution**

Energy distribution encompasses spatial and temporal regulations, with spatial distribution primarily influenced by the geographical locations of power plants, while temporal distribution pertains to fluctuations in power generation over time, especially evident in photovoltaic, hydropower, and wind power generation.

Our investigation revealed limited research focusing on the even distribution of energy within specific regions. This is predominantly attributed to being more of a managerial decision rather than a technical grid-level challenge. This observation underscores potential avenues for future research, such as the development of transmission grids, renewable energy transportation strategies, and the adaptation of blockchain data computing centers to local circumstances based on defined criteria. These criteria could be derived from geographical positioning, communication coverage, and energy consumption patterns. Numerous studies have highlighted the significant influence of geographic location on the actual implementation of blockchain technology [140]. For instance, in [51], researchers showcased how the alignment of blockchain nodes corresponded with the distribution of charging stations, underscoring the pivotal role of regional energy distribution considerations in blockchain-enabled energy markets.

In the temporal domain of energy distribution, the focus lies in making judicious adjustments to energy supply across different periods to enhance energy efficiency and user satisfaction. Blockchain technology can authenticate historical energy consumption data, providing a foundational basis for constructing temporal energy distribution models. In [141], researchers proposed a blockchain-based energy distribution model exemplifying secure parameter exchanges between various power users and enterprises to accurately model power demand, thereby reducing energy consumption and enhancing user comfort. Furthermore, Khalid et al. [85] leveraged the energy surplus of each participant as revenue, using the Shapley value method to distribute surplus energy to power-deficient entities, significantly optimizing energy utilization and addressing power deficiencies in buildings throughout the day.

**5.3 Energy trading**

Energy trading facilitates reciprocal energy production and consumption among communities when users encounter challenges in swiftly dispatching power supplies from energy producers. It is typically categorized into unilateral, centralized bilateral, and distributed bilateral markets. However, with blockchain technology revolutionizing energy trading by shedding the traditional centralized structure, this discussion shifts the focus toward unilateral markets and distributed bilateral markets. Through blockchain, transaction data integrity and traceability are ensured throughout the energy trading process, validated only when consistent across network nodes. Moreover, these transactions are seamlessly executed through automated smart contracts, guaranteeing transaction compliance and reliability.

The unilateral market comprises the buyer's and seller's markets, where one party holds a pricing advantage. In the seller's market, pricing strategies are tailored based on the level of competition among buyers. Sellers may opt for auction mechanisms or set prices in alignment with buyer budgets. For instance, Zahid et al. [142] introduced a market platform for sellers utilizing a double auction mechanism that promotes transparency and eliminates the need for intermediaries in energy management. Conversely, in scenarios with low buyer competition, sellers may resort to conventional pricing methods like cost-based, profit-based, or market-based approaches. Many theoretical models simplify by assuming unilateral market dynamics, providing pricing strategies and models under single buyer or seller scenarios.

In practice, the energy trading landscape veers towards a two-sided market paradigm [143]. In a distributed two-sided market, buyers and sellers directly engage in agreements encompassing transaction details such as object, price, time, and delivery methods. Decentralized P2P trading serves as a quintessential example of a two-sided market, essential for addressing marginal pricing, loss mitigation, voltage support, and congestion management [144]. Upholding data and user privacy is paramount in fragmented markets [145]. Market operators can deploy incentive strategies to enhance buyer and seller engagement, leveraging network externalities for increased profitability through data aggregation. Integrating technologies like smart contracts into decentralized markets presents regulatory challenges, necessitating the comprehensive utilization of blockchain layers—data, consensus, contract, and application—to devise a holistic regulatory framework for the energy market. Standardizing large-scale data transactions on multilateral platforms and enhancing pricing and transaction matching efficiency emerge as critical considerations in blockchain-driven energy markets.

## 6. Actual deployment of blockchain-based energy market

In this section, we delve into the practical implementation of blockchain-enabled energy markets in both developed and developing nations. The rationale behind this exploration stems from variations in funding availability and motivations for incorporating new information technology advancements into infrastructure across different economic landscapes. Robust financial backing and a conducive operating environment in larger nations serve as crucial prerequisites for leveraging the full potential of blockchain technology within the energy market. In [148], the authors deliberate on 140 research projects and startups leveraging blockchain in the energy sector, with the distribution across countries depicted in Fig. 7, offering insights into the existing disparities in energy blockchain investment development on a global scale. Within this discourse, our focus primarily revolves around elucidating pertinent policies, application scenarios, and corporate progress within each respective country.

### 6.1 Developed countries

As per findings presented in [149], the European Union (EU) has proposed ambitious targets to mitigate greenhouse gas emissions by 85-90% by the year 2050. This directive has instigated significant transformations within the EU's grid structure, rendering traditional large-scale power plants inadequate to meet demand requirements. Consequently, a shift towards embracing technological advancements and the industrialization paradigm of energy blockchain has been necessitated [150]. Lawrence Orsini, founder of LO3 Energy, highlighted the scarcity of global energy blockchain initiatives beyond early experimental stages, with minimal widespread adoption observed. The integration of blockchain technology in grid systems remains predominantly theoretical, with select companies in the sector grappling with the practical implementation challenges and delineating the utilization potential of blockchain within the energy landscape [151].

In light of these observations, a study outlined in [152] posited that the burgeoning energy markets emerging in the European region predominantly adhere to a designed framework. These markets operate based on historical, intra-day, and equilibrium market behaviors, boasting significant flexibility in demand-side adjustments. However, the foundational structure of these local flexible markets leans towards centralized and bilateral models, warranting close scrutiny regarding market operation intricacies and potential areas for improvement.

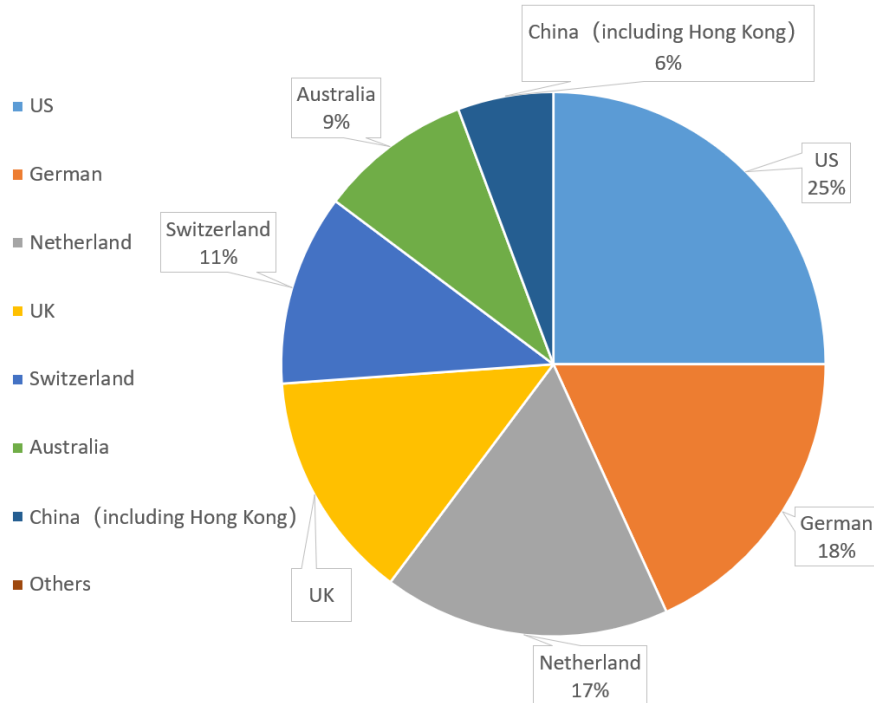

Fig 7. Percentage of energy blockchain-related platforms for 140 countries

Moreover, the United States has emerged as a prominent global player in the advancement and integration of energy blockchain technologies. A superior technological landscape and robust financial ecosystem create conducive environments for the exploration of energy blockchain research and its commercial implementation. Notably, the TransActive Grid initiative in Brooklyn, New York marks the world's inaugural energy blockchain project to transition from concept to practical application [153]. This pioneering endeavor is recognized for its decentralized facilitation of P2P transactions, enabling residents to vend surplus electricity generated by their solar panels to neighboring households via a blockchain network. Automation within this system minimizes human intervention in managing and recording transactions, streamlining operational efficiency while raising questions regarding data monitoring and consumption transparency crucial for assessing potential scalability. The implementation of such a P2P photovoltaic power project fosters electricity trade within nanoscale networks, surpassing traditional microgrid configurations by deploying photovoltaic modules on individual household rooftops, consolidating a blockchain community framework to achieve decentralized energy exchange.

Table 4. Applications of energy blockchain in developed countries

| Reference | Countries | Companies | Applications |
|---|---|---|---|
| [155] | Australia | Origin Energy | Energy trading platform |
| [156] | UK and Italy | British Petroleum (BP) | Energy trading platform |
| [157] | Japan | Tokyo Electric Power | Shared grid facility |
| [158] | Netherlands | Shell Trading International | Startup applied blockchain |
| [159] | Germany | RWD | Energy trading platform |
| [160] | Switzerland | Mercuria | Oil transaction |

Furthermore, Germany, as a prominent industrial nation, follows a structured transition towards sustainable energy practices, exemplified by the escalating installation of distributed photovoltaic generation systems in residential and commercial settings. This surge in decentralized small-scale energy production, susceptible to weather fluctuations, disrupts the conventional operations of major energy conglomerates, warranting innovative strategies to incorporate variable renewable sources into the grid network [154]. Germany emphasizes the deployment of energy storage facilities to alleviate the grid's strain due to renewable energy integration, significantly enhancing the grid's supply-demand equilibrium and operational stability. The advent of energy storage systems elevates the energy market's role, necessitating the incorporation of multi-agent consensus algorithms within blockchain technology to shape new energy market frameworks.

Furthermore, the participation of leading multinational corporations from developed nations in the energy blockchain sector is noteworthy [155-160], as detailed in Table 4, underscoring the widespread adoption of blockchain-driven energy markets and their applicability across diverse energy transactions. These instances illustrate the expansive reach of blockchain-infused energy markets, furnishing versatile solutions for various energy-related processes. The fusion of blockchain within the energy sector unveils a plethora of opportunities, streamlining functionalities such as cross-border transactions, data management, supply chain optimization, and smart contract execution.

### 6.2 Developing countries

China serves as a prominent exemplar among developing nations leading the charge in advancing energy blockchain technologies, driven by its substantial population base necessitating flexible energy supply solutions. The country's substantial investments further bolster the economic groundwork for energy blockchain development, predominantly focusing on carbon emissions tracking and carbon trading initiatives. In a significant milestone, the world's pioneering energy blockchain laboratory was established in Beijing in May 2016, paving the way for groundbreaking projects like "Demo," an energy blockchain endeavor unveiled by year-end. "Demo" introduced a digital asset known as a certified carbon reduction (CCER) - carbon ticket, essentially representing the income rights tied to carbon reduction efforts supervised by independent third-party institutions. Enterprises engaged in carbon mitigation measures receive income rights post expert evaluation and approval from forestry and relevant regulatory bodies.

Yang et al. [161] ventured into an insightful analysis of demand response dynamics within the Chinese electricity market, elucidating the impact of various economic theories on energy market pricing strategies and how these influences drive market expansion for enterprises. Noteworthy developments include the establishment of a carbon emissions trusted data certification system by Longyuan Power Company, seamlessly integrating reliable data, accounting functionalities, and carbon assets within the blockchain architecture. This system enables direct encryption of carbon transaction records on the blockchain ledger. Concurrently, Far Light Software Enterprise spearheads pioneering products such as blockchain-enabled storage solutions for the settlement of distributed energy transactions, underscoring their commitment to supporting blockchain-powered "green power traceability" technologies for the upcoming 2022 Beijing Winter Olympics. The widespread blockchain integration in diverse sectors in China, as highlighted by Yang et al. [162], emphasizes the potential of incorporating blockchain within energy market regulations, enabling participants to verify identities, securely exchange sensitive information, bolster operational efficiency, and contribute towards reducing carbon emissions.

In the context of other developing nations, there remains a pressing demand for increased investment in blockchain-based energy markets to facilitate the modernization and enhancement of local energy systems. Notably, in regions such as relatively impoverished Africa, initiatives like Sinan Energy's carbon credit trading platform [163] are designed to meticulously document and validate carbon emissions from power facilities, subsequently engaging in carbon trading activities on the global stage. However, energy trading practices within African territories suffer from opacity and fragmented pricing mechanisms, leading to protracted transaction cycles with inflated intermediary costs. Consequently, leveraging blockchain technology to support the crowdfunding of clean energy projects in economically disadvantaged regions of Africa has gained substantial traction. Additionally, blockchain integration holds the promise of rendering energy production, storage, transmission, trading, utilization, and management processes more intelligent and environmentally sustainable, expediting the implementation of initiatives like smart grids and energy Internet infrastructure across Africa.

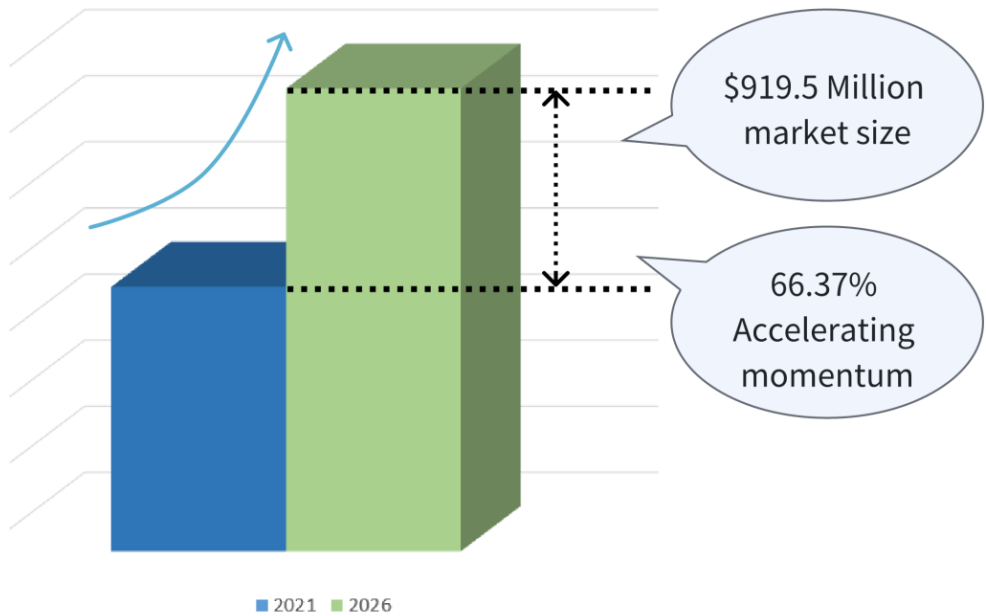

Fig 8. The market value of energy blockchain.

Moreover, the Philippines grapples with inadequate infrastructure, particularly in the power grid sector, exacerbating recurrent power outages. Consequently, the nation is exploring the deployment of microgrids to enhance the reliability of electricity provision [164]. The inherently decentralized nature of blockchain technology can facilitate diverse microgrid users in conducting inter-regional point-to-point power transactions to ensure a harmonious balance between power supply and demand.

Drawing upon the preceding analysis, developed nations must navigate and mitigate potential risks associated with blockchain adoption, encompassing aspects like information security and financial integrity. Simultaneously, these nations must adopt a proactive stance towards embracing the blockchain-enabled energy market at a global scale, effecting responsive measures to address challenges related to energy conservation and emission reduction. Correspondingly, developing countries are advised to pursue gradual advancements while upholding stability. In an epoch characterized by burgeoning data volumes and depleting energy reservoirs, advancing economic investments and research efforts in energy blockchain tailored to each nation's context are crucial in propelling the energy system toward substantial qualitative transformation and progression.

A comprehensive market analysis by the esteemed global technology research and consultancy firm, Technavio, forecasts significant growth in the blockchain technology segment within the energy market, projecting a remarkable increase of $919.5 million between 2021 and 2026 [175]. During this forecast period, a robust growth rate of 66.37% is anticipated, as illustrated in Fig. 8. Against the backdrop of renewable energy emerging as a pivotal player in energy trading, the blockchain-enabled energy market catalyzes propelling the global transition towards a low-carbon future.

## 7. Challenges

Amidst the profound shifts in the global climate landscape, the evolution of the energy market continues on an upward trajectory. An increasing array of nations are turning to blockchain technology as a transformative tool for revolutionizing energy markets. Notable applications include P2P pricing models, peak trend analyses, and smart contracts facilitating enhanced energy consumption management with a central aim of mitigating and potentially offsetting carbon emissions. Nevertheless, the pervasive emergence of numerous challenges and constraints prompts a critical examination of strategies aimed at addressing and surmounting these obstacles within the blockchain-driven energy market. This imperative focus on resolving challenges within the blockchain-enabled energy market underscores a pivotal area of concern warranting dedicated attention [166].

### 7.1 Blockchain scalability

The expanding energy market exerts heightened pressure on blockchain scalability, as evidenced by challenges in achieving extensive information interconnection and sharing. Within the energy market framework, frequent activities such as energy generation, scheduling, and trading result in the generation of blocks that necessitate incorporation into a chain structure [167]. However, the operational efficiency of blockchain networks in processing these blocks remains constrained, potentially leading to network congestion when a substantial volume of blocks awaits validation [168]. This places a significant premium on enhancing the scalability of blockchain technology. In instances where the operational speed of the blockchain fails to meet the demands of power plants and users, it can precipitate disruptions in power supply, a consequence more severe compared to other network scenarios as it not only disrupts transactions but also compromises the security and stability of the power grid.

Consequently, numerous researchers have proposed innovative approaches to overcome the limitations inherent in traditional blockchain systems. For instance, Gramoli et al. [169] advocate for leveraging a Directed Acyclic Graph (DAG) to safeguard user and node transaction data, potentially supplanting the conventional single-chain architecture in blockchain systems and bolstering block writing efficiency [170]. Additionally, researchers are exploring sharding technology as a means to alleviate the burden on blockchain-wide consensus mechanisms. This approach involves partitioning network nodes into distinct clusters wherein each cluster independently achieves consensus, thereby enhancing the parallelism of consensus processes [171]. Nonetheless, these solutions are still in the nascent stages of exploration. Notably, the prevailing blockchain technology exhibits limitations in terms of data handling capacity despite its widespread popularity in the market.

**7.2 Energy consumption of blockchain**

As discussed earlier, the dynamic operations within the energy market, involving energy allocation, trading, and related activities, result in the generation of a substantial volume of blocks that necessitate processing within the blockchain system. Notably, the operation of blockchain technology itself is known to be energy-intensive [140], [172-173]. Therefore, the heightened frequency of blocks generated by the frenetic energy market activities contributes to a heightened energy consumption and operational burden on the blockchain system.

The majority of users engaging with blockchain systems contend with limited bandwidth and computer processing capabilities, underscoring a significant challenge confronting blockchain system designers to minimize user network energy consumption while maximizing network performance. The foundational premise of blockchain-driven energy markets was rooted in reducing energy consumption. Hence, any substantial energy loss attributable to blockchain operation would diverge from this core principle and incur additional costs to uphold system security standards. Concerns regarding the sustainability and carbon emissions impact of the prevailing Bitcoin ecosystem have intensified over time, prompting a critical examination of strategies to curtail electricity consumption for long-term viability, as noted in [174]. In the realm of power generation, the primary application domain centers around distributed energy sales facilitated by microgrid-based blockchain systems.

Notably, inefficiencies in power transportation lead to electricity losses, with transaction costs escalating with distance. For distributed photovoltaic power generation systems operating at lower voltage levels, electricity transportation and trading are ideally executed in proximity to the energy generation location, restricting distributed energy utilization to small-scale applications. The aforementioned challenges significantly constrain potential blockchain application scenarios and operational scope, underscoring emergent hurdles within the domain.

**7.3 Storage overhead and response speed of blockchain**

The concerns surrounding the storage overhead and response speed of blockchain form another critical area of exploration and analysis within our research domain [175-176].

The distributed storage architecture inherent in blockchain systems leads to substantial storage redundancy [177]. The heightened frequency of energy transactions in the energy market generates a considerable volume of transactional data that necessitates storage, thereby exacerbating the storage burden on the blockchain system.

Furthermore, there remains ample room for enhancing the response speed of blockchain systems compared to well-established bank transaction mechanisms [176]. Notably, prominent blockchain platforms like Bitcoin and Ethereum exhibit limitations in processing financial transactions at scale, often resulting in transaction delays. Even the swiftest private chains are susceptible to such constraints. The consensus mechanisms integral to blockchain operations necessitate extensive node communication and data exchange, contributing to delayed response times.

Presently, the modest scale of existing energy blockchain projects implies relatively straightforward application scenarios with manageable complexities. However, with the burgeoning expansion of blockchain network sizes, challenges pertaining to storage overhead and response speed escalate in significance. This becomes particularly pronounced when envisioning the application of blockchain in the ultra-short-term optimal scheduling of multi-energy integrated systems in the future [178], where potential limitations in information processing speeds may not arise to meet operational demands.

### 7.4 User Concern about blockchain

As a pivotal indicator of the blockchain industry's pulse, the ebbs and flows in Bitcoin's price trajectory serve as a reflection of market sentiments and ambivalence towards blockchain technology in recent years [179], illustrated in Fig. 9 [180].

From this vantage point, the enduring enthusiasm for blockchain technology warrants thoughtful consideration, especially concerning its integration into the energy market. Everyday consumers, grappling with constraints imposed by price regulations in retail electricity and the complexities of redundant energy trading processes, may harbor reservations about the potential for substantial and sustained returns. Rather than perceiving blockchain-based initiatives as lucrative opportunities, concerns may arise about the possibility of escalating electricity prices, fostering apprehension among consumers. In the absence of substantial demand and tangible benefits, consumer engagement in blockchain-driven energy markets may face significant hurdles. Moreover, the prevalence of blockchain technology in the energy sector pales in comparison to its widespread utilization in flourishing domains like financial blockchain applications. Consequently, policymakers are encouraged to proactively advocate for increased investments in energy blockchain solutions while conceptualizing models that offer mutual advantages for power plants and consumers.

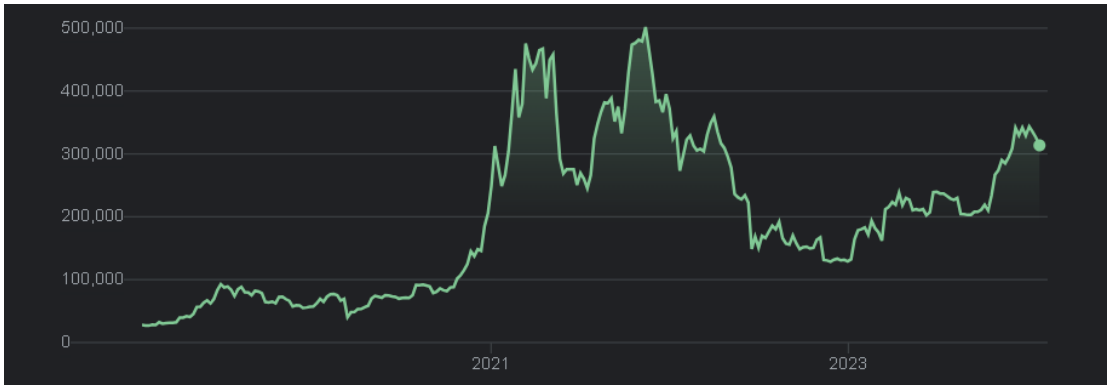

Fig 9. Fluctuating Bitcoin prices [180].

An effective approach to garnering user interest in blockchain entails emphasizing the paramount importance of personal privacy in energy transaction processes. Users are understandably averse to scenarios where their privacy is compromised and fairness is jeopardized. For power producers, inadvertent privacy breaches could potentially expose proprietary trade secrets, paving the way for unfair competition. On the other hand, for power consumers, privacy breaches may unveil sensitive personal information, ultimately leading to adverse incidents like theft [181].

8. Promising aspects

This section delves into an examination of the prospective development trends within the blockchain-based energy market. Our intention is to conduct a comprehensive exploration of this subject, commencing with a thorough analysis encompassing both technological strategies and policy considerations.

**8.1 Technical strategy**

As delineated in the preceding section, the challenges faced by the blockchain-powered energy market, including high energy consumption, elevated storage overhead, and sluggish response rates, predominantly stem from deficiencies in blockchain scalability. Consequently, scalability emerges as a pivotal bottleneck necessitating targeted interventions for the advancement of the blockchain-based energy market.

One viable solution to address scalability concerns involves the adoption of cross-chain technology. This innovative approach fosters blockchain interoperability through the implementation of cross-chain message transfer protocols, empowering the creation of cross-chain decentralized applications (DApps) capable of deploying smart contracts across diverse blockchains [182]. Notably, such protocols enhance blockchain interoperability and performance, offering substantial benefits within the energy market domain. Leveraging cross-chain technology enables the tailored design of specific blockchains catering to the unique requirements of the energy industry, facilitating seamless transaction and data exchange between blockchains to effectively tackle scalability and transaction speed challenges [15].

Alternatively, the implementation of sharding presents another promising avenue to tackle scalability hurdles by facilitating horizontal scaling. Sharding involves the segmentation of network nodes into distinct clusters known as shards based on predefined criteria. These shards can diligently process transactions in parallel upon transaction initiation within the blockchain network, significantly bolstering overall processing speed and efficiency [172]. Moreover, adopting a sharding approach minimizes the need for every node to download the entire transaction data set, optimizing storage resource utilization [177]. Illustrative examples of sharding implementations include OmniLedger [183], Atomix [184], among others. Despite the efficacies of sharding protocols, potential limitations linked to an increased volume of cross-shard transactions emphasizing the need for judicious transaction allocation methods to forestall throughput degradation underscore the imperative need for further exploration and refinement in this area [185].

Furthermore, scalability challenges can also be partially mitigated through adjustments in block size, modifications to block structures, decoupling block generation, and conceptualizing innovative chain structures such as Direct Acyclic Graphs (DAG), among others [186]. While these technologies have yet to achieve widespread adoption and exploration within the energy market domain, they offer promising avenues for future research and development efforts to enhance scalability and operational efficiency within the blockchain-based energy market landscape.

**8.2 Government policy**

In principle, blockchain has the potential to revolutionize every facet of the energy industry value chain. However, as elaborated in the preceding section, the transition of energy industry enterprises to blockchain technology remains premature due to prevalent technical challenges. Hence, the establishment of a robust blockchain-based energy market necessitates immediate policy cultivation and guidance from pertinent government entities.

In the context of developing countries, the adoption of blockchain technology poses greater challenges and offers fewer incentives. The constrained per capita GDP levels render it infeasible for these nations to meet their fundamental energy supply and demand needs, impeding the acceptance of blockchain technology that mandates substantial capital investment. Conversely, in developed countries where scientific and technological innovation thrives, the maturity of blockchain implementation is contingent upon policy backing. Given the stringent regulatory oversight characteristic of the energy sector as a foundational industry, projects in this domain mandate policy approvals or qualification acquisitions before implementation.

Hence, national policy directives wield a profound influence on the convergence of blockchain technology and the energy sector. For instance, in [187], the authors assert that a judicious regulatory framework and trading regulations could empower the UK electricity market to transition towards a more flexible, low-carbon energy system. Furthermore, as illustrated in Fig. 10, shifts in blockchain demand and interest are contingent upon regulatory policies. In essence, sound and well-crafted national policies serve as the cornerstone for solidifying the blockchain industry's foothold and catalyzing the expansion of the energy market. This can be achieved through the enactment of laws and industry standards that delineate the role of blockchain in energy markets, while government-led demonstration initiatives can vividly showcase the tangible benefits for both energy consumers and power plants.

9. **Conclusion**

This survey critically examines the prevailing scenario and emerging trends within the blockchain-based energy market. Initially, a meticulous review of influential literature in the domain was conducted to establish foundational insights. Subsequently, an exploration is undertaken to ascertain the rationale behind blockchain's applicability in the energy market, dissecting the technological facets, practical applications, and operational models facilitated by blockchain integration. A detailed analysis of the challenges inherent to blockchain-based energy markets is then presented, offering a comprehensive overview of the impediments faced by the sector. Lastly, a forward-looking perspective is outlined concerning the industry, encompassing technological strategies and governmental policies.

In essence, this study delves into not only the enabling technologies for blockchain-driven energy markets but also sheds light on their practical implementation modalities. A synthesis of real-world deployments is encapsulated within this analysis. The intent behind this research endeavor is to furnish theoretical guidance for researchers in the field and furnish pertinent market analysis strategies for relevant stakeholder entities.